\documentclass[iop,revtex4]{emulateapj}
\usepackage{psfig,natbib}
\newcommand{\figdir}{./}
\newcommand{\mfigur}[3]
    {\begin{figure}[hbt]
        \centerline{\psfig{figure=\figdir/#1.ps,height=#2}}
        \caption{\label{#1}#3}
    \end{figure}}
\def\note #1]{{\bf #1]}}
\providecommand{\msize}{\normalsize}
\def\aatab#1#2#3#4#5#6{\ifthenelse{\equal{*}{#1}}
{\begin{table*}[htbp]\msize\caption[]{\label{#2} #3}
  \begin{flushleft}
    \begin{tabular}{#4}
      \hline\noalign{\smallskip} #5
          \noalign{\smallskip} \hline \noalign{\smallskip} #6
          \noalign{\smallskip} \hline
        \end{tabular}
  \end{flushleft}
\end{table*}}
{\begin{table}[htbp]\msize\caption[]{\label{#1} #2}
  \begin{flushleft}
    \begin{tabular}{#3}
      \hline\noalign{\smallskip} #4
      \noalign{\smallskip} \hline \noalign{\smallskip} #5
      \noalign{\smallskip} \hline
    \end{tabular}
  \end{flushleft}
\end{table}}}
\newcommand{\oh}[1]{\hidewidth #1\hidewidth}
\newcommand{\be}[1]{\begin{equation}\label{#1}}
\newcommand{\ee}{\end{equation}}
\newcommand{\bea}[1]{\begin{eqnarray}\label{#1}}
\newcommand{\eea}{\end{eqnarray}}
\newcommand{\mideqn}{\nonumber\\[-2\jot] &{\vbox{\hsize=0pt}}&
    \nonumber\\[-\baselineskip] &{\vbox{\hsize=0em}}&\\[-.5\baselineskip]}
\newcommand{\mfrac}[2]{{\textstyle{#1\over #2}}}
\newcommand{\dd} [2]{{{{\rm d}{#1}\over{\rm d}{#2}}}}

\newcommand{\dxdy}[2]{\frac{\partial #1}{\partial #2}}
\newcommand{\stimes}{\times}
\newcommand{\p}{\phantom{2}}
\newcommand{\eg}  {e.g.}            
\newcommand{\ie}  {i.e.}            

\begin{document}
    \title{A Grid of 3D Stellar Atmosphere Models of Solar Metallicity:\\
           I.\ General Properties, Granulation and Atmospheric Expansion}
    \shorttitle{Grid of 3D Atmosphere Models}

    \author{Regner Trampedach}
    \affil{JILA, University of Colorado and National Institute of Standards
           and Technology, 440 UCB, Boulder, CO 80309, USA}
    \and    \author{Martin Asplund and Remo Collet\altaffilmark{1,2}}
    \affil{Research School of Astronomy and Astrophysics,
           Mt.\ Stromlo Observatory, Cotter Road, Weston ACT 2611, Australia}
	\and    \author{{\AA}ke Nordlund\altaffilmark{2}}
    \affil{Astronomical Observatory/Niels Bohr Institute, Juliane Maries Vej 30,
           DK--2100 Copenhagen {\O}, Denmark}
    \and    \author{Robert F. Stein}
    \affil{Department of Physics and Astronomy, Michigan State University
           East Lansing, MI 48824, USA}

    \altaffiltext{1}{Astronomical Observatory/Niels Bohr Institute,
                    Juliane Maries Vej 30, DK--2100 Copenhagen {\O}, Denmark}
    \altaffiltext{2}{Centre for Star and Planet Formation,
                    Natural History Museum of Denmark, University of Copenhagen,
                    {\O}ster Voldgade 5--7, 1350 Copenhagen, Denmark}


    \begin{abstract}
Present grids of stellar atmosphere models are the workhorses in
interpreting stellar observations, and determining their fundamental
parameters.
These models rely on greatly simplified models of convection, however,
lending less predictive power to such models of late type stars.

We present
a grid of improved and more reliable stellar atmosphere models of
late type stars, based on deep, 3D, convective, stellar atmosphere
simulations. This grid is to be used in general for interpreting
observations, and improve stellar and asteroseismic modeling.
        
We solve the Navier Stokes equations in 3D and concurrent with the
radiative transfer equation, for
a range of atmospheric parameters, covering most of stellar
evolution with convection at the surface.
We emphasize use of the best available atomic physics for quantitative
predictions and comparisons with observations.
        
We present granulation size, convective expansion of the acoustic
cavity, asymptotic adiabat, as function of atmospheric parameters. These
and other results are also available in electronic form.
    \end{abstract}


    \maketitle

	\section{Introduction}
	\label{intro}
We present a homogeneous grid of three dimensional (3D) convective
stellar atmosphere models, with applications to a wide range of astronomical
inquiries. Previews of the grid have previously been given by
\citet{trampedach:Gough2007,trampeda:mixlength} with recent asteroseismic
applications by \citet{savita:RG_granulation,savita:Kpl22sols} and
\citet{bonaca:KeplerAlpha}. The simulations have been computed with the
\citet{stein:solar-granI}-code, but with updated atomic physics as described
in more detail in Sect.\ \ref{phys}.
The solar abundance determination by \citet{AGS05} was based
on a solar simulation identical to ours in all respects, except for a slightly
higher horizontal resolution.

The first systematic analysis of convection in stellar atmospheres, through 3D
simulations, was performed in the series of papers by
\citet{stell-gran3,stell-gran4}. They were interested in the 3D structure of
granulation and the effects on spectral lines. These simulations were computed
with an anelastic precursor to the present, fully compressible
\citet{stein:solar-granI}-code.

Currently an independent grid, the CIFIST grid \citep{ludwig:CIFIST-grid}, is
being assembled using the CO$^5$BOLD code \citep{CO5BOLD,freytag:CO5BOLD} to
compute 3D simulations of late-type stars, including sub-solar metallicities and
M-type stars.

A new grid of 3D convection simulations using the Stagger-code is
nearing completion
\citep{remo:StaggerGrid,magic:stagger-grid}, and will encompass both sub- and
super-solar metallicities. We are truly entering the age of 3D convective
atmosphere simulations, and in combination with recent and soon to be realized
observational advances, we are sure to learn much about the inner workings of
stars in the coming decades.

This paper is an attempt at
publishing results of 3D convection simulations in a format similar to what
astronomers have been using for the last half a century: a grid in the
atmospheric parameters
effective temperature, $T_{\rm eff}$, surface gravity, $g$, and metallicity,
[Fe/H]. This grid, however, is only for solar metallicity, [Fe/H$]=0$.

A major motivation for this effort, has been the advent of asteroseismology
and the significant leap in stellar observations provided by the MOST
\citep{walker:MOST}, COROT \citep{baglin:COROT-IAUColl185} and Kepler
\citep{borucki:Kepler} missions, and even by the star tracker on the failed
infra-red satellite, WIRE \citep[e.g.,][]{bruntt:ProcyWIRE}. These observations,
with their highly stable, precision photometry and long time-lines,
can reveal a wealth of detail about \emph{field stars} that could previously
only be considered for the Sun and a handful of other stars. The sheer volume
of Kepler observations also means that a large range of stellar conditions are
well sampled and meaningful statistics can be developed.
The much anticipated launch in 2013 of the Gaia mission
\citep{perryman:GAIA,prusti:GaiaPromise} will greatly improve our knowledge
of stellar distances, including for most Kepler targets, providing strong and
independent constraints on absolute luminosities and radii.
These observational advances demand reciprocal advances on the modeling side,
and the present grid of convection simulations is our contribution to such an
advance.

The present work is an alternative to previously published grids of 1D stellar
atmosphere models. Some of the 1D atmosphere grids that have found widespread
use have been constructed by
\citet[MARCS]{MARCS-2}, \citet[ATLAS9]{kurucz:newATLAS9} and 
\citet[NextGen-grid of PHOENIX models]{hauschildt:nextgen-atm,hauschildt:nextgen-grid2}.
These models vary in atomic and molecular line databases, continuum opacities,
equation of state, whether they employ plane-parallel or spherical geometry
and to what extent non-LTE effects are included. In all these cases radiative
transfer is the focus and it is solved in impressive detail and with high
accuracy. For late type stars, they also all share the same short-coming: The
analytical formulation of convection in the form of the widely used mixing
length theory pioneered by \citet[MLT]{boehm:mlt} or the more recent variant
by \citet{canuto-mazzitelli:conv,canuto-mazzitelli:conv-improv}.
In contrast, the present atmosphere grid is based on explicitly evolving the
Navier Stoke's equations, as detailed in Sect.\ \ref{phys}, with the
hydrodynamics being directly coupled with realistic radiative transfer
(see Sect.\ \ref{radtransf}).

The lay-out of our grid of 37 simulations, is described in Sect.\ \ref{thegrid},
where the issues of interpolating in the grid, and methods for averaging of
the simulations are also discussed. General properties of convection in the
simulations are discussed in Sect.\ \ref{GenConv}, and contrasted with
the analytical MLT approach. Here we also touch on the effect of 3D convection
on the frequencies of p-modes.

The surface manifestation of convection on the Sun, has been observed for
more than two centuries \citep{herschel:SunGran}, with the term \emph{granules}
coined by \citet{dawes:SunGranules}. Granulation on other stars has been rather
elusive, though, with a tentative observation on Betelgeuse ($\alpha$\,Ori)
by \citet{lim:BetelgeuseGran}. Our simulations reproduce the quiet Sun
granulation, both with respect to size, contrast and general shapes, giving us
confidence in our predictions for other stellar parameters, as presented in
Sect.\ \ref{varTg}. Recently, the unprecedented time-line, stability and
precision of the Kepler observations, has allowed direct observations of the
effect of granulation on integrated star light from red giants
\citep{savita:RG_granulation}. These observations were analyzed and compared
with the present grid of simulations, and the agreement is generally good,
although unresolved issues remain. A more detailed analysis, also including
main sequence (MS) stars, will be carried out in the near future.

We round off with conclusions in Sect.\ \ref{conclusion}.


	\section{The 3D Convection Simulations}
	\label{phys}

The simulations evolve the fully compressible Navier Stoke's equations for
mass, momentum and energy conservation
\bea{NavStok}
    {\partial\ln\varrho\over \partial t} &=& -\vec{u}\cdot\nabla\ln\varrho
                                           - \nabla\cdot\vec{u} \nonumber\\
    {\partial{\vec u}\over \partial t} &=&
                - \vec{u}\cdot\nabla\vec{u}
                + \vec{g} - {P_{\rm g}\over\varrho}\nabla\ln P_{\rm g}
                + \frac{1}{\varrho}\nabla\cdot\vec{\sigma}\\
    {\partial \varepsilon\over \partial t} &=& 
                - \vec{u}\cdot\nabla\varepsilon
                - {P_{\rm g}\over\varrho}\nabla\cdot\vec{u}
                + Q_{\rm rad} + Q_{\rm visc}\ ,\nonumber\label{Econs}
\eea
where $\varrho$ is the density, $P_{\rm g}$ the gas pressure, $\varepsilon$ the
internal energy per unit mass, $\vec{u}$ the velocity field and $\vec{g}$ is the
gravity, assumed to be constant with depth. This version of the equations,
differ from the normal version, by having been divided by $\varrho$ to
pre-condition the equations for the very large density gradients in stellar
surface layers.

The \citet{stein:solar-granI}-code used here is a non-staggered, finite
difference code.
Derivatives and interpolations are evaluated from cubic splines and time is
advanced with a third order leapfrog (predictor/corrector) scheme
\citep{hyman:LeapFrog,aake:comp-phys}.

As all higher-order (higher than linear) order schemes are unstable artificial
diffusion is needed for stabilizing the solution. We employ a quenched
hyper-diffusion scheme which efficiently concentrates the diffusion at sharp
changes in variables, leaving the smoother parts of the solution unaffected.
Hyper-diffusion means a fourth-order term is added to the normal second-order
(Laplacian) term, and the quenching consists of ensuring the fourth-order term
everywhere has the same sign as the second order term, to not excite more
ringing. This is explained in more detail by \citet{stein:solar-granI}. This
does not constitute a sub-grid scale model, as we have not specified any
preconceptions of what should happen below the grid-scale of our simulations.
That in turn implies that those scales do not contribute any significant net
fluxes, {i.e.}, motions on that scale are considered isotropic turbulence.

This
diffusion by the viscous stress tensor $\vec{\sigma}$, gives rise to the
dissipation, $Q_{\rm visc}=\sum_{ij}\sigma_{ij}\partial u_i/\partial r_j$, in Eq.\ (\ref{NavStok}).
The radiative heating, $Q_{\rm rad}$, is dealt with in more detail in
Sect.~\ref{radtransf}.

Each of the simulations has been performed on a $150\stimes 150\stimes 82$
rectangular grid with equidistant and cyclic horizontal grid and a
vertical (smooth)
grid which is optimized to capture the large temperature gradients in
the photosphere. The simulations cover about 13\,pressure scale-heights
vertically, with half of those below and half above the photosphere. On a
Rosseland optical depth scale, this corresponds to
$\log_{10}\tau_{\rm Ross}=[-4.5;\,$5--8$]$. This is the region we will
refer to as \emph{stellar surface layers}.
The effects of numerical resolution on stratification and velocity fields
have been explored by \citet{asplund:num-res}, on which we based our choice
of resolution.

Both the top and bottom boundaries are penetrable with outflows leaving
undisturbed. To approximate the effect of the convection zone below the
simulation domain, \emph{inflows} at the bottom boundary are evolved
towards a
constant entropy, hydrostatic equilibrium and a vertical velocity that
balances the mass flux of the unaltered outflows. The inflows are evolved
towards this ideal solution, on time-scales of half the minimum 
flow-crossing time at the bottom, which means the inflow profiles are
not entirely flat. The entropy is in practice set by values of density
and internal energy. The total flux of the simulation, $\sigma T^4_{\rm eff}$,
is solely due to the supplied entropy of the inflows---no artificial fluxes
are imposed anywhere. Effects of the bottom boundary are imperceptible in
temperature, pressure and density, but recognized in entropy, superadiabatic
gradient and velocities, for which they affect the bottom 5--10\,grid-points,
covering 0.6--1.0\,pressure scale-heights.
The top boundary is a fiducial
layer, located three times further away from the first grid point in, than
the next. This together with the exponentially declining density,
ensures only little mass is present there, and minimizes its
influence on the simulation. Velocities and energies from the first grid
point are simply copied to the fiducial layer, and the density is extrapolated
hydrostatically.

\subsection{Atomic Physics}

We employ the so-called Mihalas-Hummer-D{\"a}ppen (MHD) equation of state (EOS)
\citep{mhd1,mhd2,mhd3} which includes explicit dissociation, ionization and
excitation of all states of all species of all included elements. The
only two molecular species included are H$_2$ and H$_2^+$, however. We have
custom calculated tables of this EOS for the chemical mixture listed in Table
\ref{abundlist}, which includes 15 elements as opposed to the previously
published 6 element mixtures. As the dependent variables of the Navier
Stoke's equations are $\varrho$ and $\varepsilon$, we invert the EOS table
to one in $(\ln\varrho, \varepsilon)$. The entropy is evaluated by
integration of the table
\be{Sintegrate}
    S = S_0 + \int \frac{1}{T}\left({\rm d}\varepsilon -
                    \frac{p_{\rm gas}}{\varrho}{\rm d}\ln\varrho\right)\ ,
\ee
where we first integrate along iso-chores of the table and then along the
internal energy axis. The integration constant, $S_0$, is (arbitrarily) chosen
so as to have zero-point at $\varrho = \sqrt{10}\stimes 10^{-7}$\,g\,cm$^{-3}$
and $\varepsilon = 2.5\stimes 10^{12}$\,erg\,cm$^{-3}$.

The bound-free and free-free opacities, as well as scattering cross-sections
are calculated using the MARCS package of codes \citep{b.gus}, which is the
engine behind the MARCS stellar atmosphere models \citep{gus:modgrid,MARCS-2}.
The data-tables, however, are updated as follows:

We have included the absorption by H$^-$ bound-free (bf)
\citep{broad-rein:H-,wish:H-}, free-free (ff) \citep{bell-berr:H-ff},
H$_2^+$ bf+ff \citep{stancil:H2+}, H$_2^-$ ff \citep{bell:H2}, and the
photo-dissociation of OH and CH \citep{kur:OH+CH}.
The most important levels of \ion{He}{1}, \ion{C}{1}, \ion{N}{1}, \ion{O}{1},
\ion{Na}{1}, \ion{Mg}{1}, \ion{Mg}{2}, \ion{Al}{1}, \ion{Si}{1}, \ion{Ca}{1},
\ion{Ca}{2} and \ion{Fe}{1} were included as simple analytical fits as
listed by \citet{mathisen1}. Bound-free absorption by \ion{O}{1} and \ion{C}{2}
were adopted from the MARCS \citep{MARCS-2} implementation of OP opacities
\citep{butler:OP-OI-bf,OP:CII}. Most of these changes were described in more
detail by \citet{trampedach:thesis}.

Rayleigh scattering is included as fits to the results by \citet{gavrila:Hray}
for \ion{H}{1}, \citet{Lang:Heray} for \ion{He}{1} and by
\citet{Vict-Dalg:H2ray} for H$_2$. The usual Thomson scattering by free
electrons is also included.

The line opacity is supplied by the opacity distribution functions (ODFs) of
\citet{kur:line-data,kur:missolar}, which include 58\,million lines from the
first nine ions of all elements through Ni and the
diatomic molecules H$_2$, CO, CH, CN, C$_2$, NH, OH, MgH, SiH, SiO and TiO.

\subsection{Chemical Abundances}
\label{abunds}

The Solar abundances we use, are listed in Table \ref{abundlist} as logarithmic
abundances normalized to $A({\rm H})=12.00$. In our adaptation of atomic physics, the
abundances enter in three independent places; In the EOS calculation,
the bf- and ff-opacities and in the line opacities supplied by the ODFs.
Since we have calculated our own tables of the MHD EOS and perform our own
calculation of bf- and ff-opacities using the MARCS package, we are free
to choose the abundances for those two parts -- only restricted in the choice 
of elements by the availability of the necessary atomic physics.

\begin{deluxetable}{rlrrlr}
\tablecaption{Logarithmic abundances normalized to $A({\rm H})=12.$
    \label{abundlist}}
\tablewidth{210pt}
\tablehead{
\colhead{$Z$}& \colhead{\strut Elem.} & \colhead{Abund.}\hskip -.6em & 
\colhead{~~~~~~~$Z$}& \colhead{\strut Elem.}& \colhead{Abund.}\hskip -.6em}
\startdata
 1 & H  &  12.00 & 14 & Si &   7.51 \\
 2 & He &  10.92 & 16 & S  &   7.17 \\
 6 & C  &   8.52 & 18 & Ar &   6.52 \\
 7 & N  &   8.01 & 19 & K  &   5.08 \\
 8 & O  &   8.89 & 20 & Ca &   6.32 \\
10 & Ne &   8.05 & 24 & Cr &   5.63 \\
11 & Na &   6.29 & 26 & Fe &   7.50 \\
12 & Mg &   7.54 & 28 & Ni &   6.21 \\
13 & Al &   6.43 & & &
\enddata
\end{deluxetable}

For the ODFs, however, we have less control over the assumed composition,
and we have therefore largely
adopted the abundances prescribed by the available ODFs at the time of starting
the simulation grid.
\citet{kur:line-data,kur:missolar} has made ODF tables available for
the abundances by \citet[AG89 hereafter]{AG89} for a
range of metallicities, but also for a He-free mixture and a few tables for
lower Fe abundances. This enabled us to perform interpolations to the
helioseismically determined Solar He abundance of $A({\rm He})=10.92$
(down from the classic value of 
$A({\rm He}=11.00$) \citet{basu-antia:Y,basu:Rsun-err} and the more modern Fe abundance
of 7.50 \citep{GS98,AGSS2009} (down from 7.67). These interpolations resulted
in the abundances listed in Table \ref{abundlist}.
The differences between ours and the recent abundances based on solar
convection simulations, \citep[AGSS'09]{AGSS2009}, are rather small and not
systematic, except for the decreased C, N and O abundances. These three
elements, however, have little effect on atmospheric structure, as they supply
only little line  or continuum opacity, and affect the EOS minimally compared
to the ionization of hydrogen (they have ionization energies similar to H). The
only significant influence is through the molecules of H, C, N and O. Solar
ATLAS9 models by \citet[only on-line]{kurucz:newATLAS9} show very small
differences in structure, though, between GS'98 and AGS'05: The latter is 20\,K
warmer at $\log\tau_{\rm Ross}=-6.0$ decreasing linearly to a few Kelvins
around the photosphere and is about 30\,K cooler below
$\log\tau_{\rm Ross}\sim 1$. AGSS'09 has C, N and O abundances that are closer
to the ''classic'' GS'98 values, than does AGS'05, and we expect the above
differences to be even smaller.

For these reasons we did not
feel compelled to recompute the, at that time, mostly finished grid of
simulations to adopt the new abundances.

    \subsection{Radiative Transfer}
	\label{radtransf}

The radiative transfer in the atmosphere, has been described in detail by
\citet{aake:numsim1,stell-gran3,bob:Tuebingen}. The forward solution is
performed on long
characteristics, and the original \citet{feautrier:method}-technique
\citep[See also][]{mihalas:stel-atm}, has been modified to give more
accurate solutions in the optical deep
layers \cite{aake:numsim1}. The angle to the vertical of an outgoing
characteristic
(or ray) is $\mu=\cos\theta$ and the azimuthal angle is $\varphi$.

The angular integrations are performed with $N_\varphi=4$ $\varphi$-angles
and $N_\mu=2$ $\mu$-angles with $\mu$-points and -weights determined by
Radau quadrature \citep{radau:quadrature,abramowitz}. This method ensures the
highest accuracy when one end-point (the vertical) is to be included.
For $N_\mu=2$ this method gives $\theta=0^\circ$ and $70^\circ$ with weights
$\mfrac{1}{4}$ and $\mfrac{3}{4}$. The $\phi$-angles are rotated for each
time-step to avoid developing preferred directions.
To perform radiative transfer on slanted rays, the whole simulation cubes
are rotated and tilted by rotating and shifting each horizontal plane sideways,
exploiting the periodic boundaries. This is performed with bi-cubic spline
interpolation. The number of grid-points in the inclined cubes is the same
as in the normal, vertical cube, and the computational cost is therefore also
the same for the vertical as for each $(\mu,\phi)$ inclined calculation of
the radiative transfer.


The effects of spectral lines are included through the method of opacity binning.
This consists of grouping wavelengths together based on their opacity strength,
or more precisely, the Rosseland optical depth, $\tau_{\rm Ross}$, at which 
the monochromatic optical depth is $\tau_\lambda=1$. We use four bins, with
each bin spanning a decade in opacity strength. The source function used for
each bin is simply the Planck function, $B_\lambda(T)$, summed up over the
wavelengths, $\lambda_j$, in each bin, $i$,
\be{Bi}
    B_iw_i = \sum_{j(i)} B_{\lambda_j}w_{\lambda_j}\ ,\quad
    w_i    = \sum_{j(i)} w_{\lambda_j}\ .
\ee
The weights, $w_{\lambda_j}$, for each wavelength is simply the wavelength
width of each ODF, multiplied by the weight of each bin in the ODF.
This method of opacity binning ensures that the radiative cooling and heating
in the photosphere of strong lines is included as well as that of the continuum.
The radiative heating per volume is
\bea{Qrad}
    Q_{\rm rad} &=& 4\pi\varrho\int_\lambda
                    \kappa_\lambda(J_\lambda-S_\lambda){\rm d}\lambda\\
            &\simeq& 4\pi\varrho\kappa_{\rm Ross}\sum_i x_i(J_i-B_i)w_i\ ,
\eea
where $J_\lambda$ is the zero'th angular moment of the specific intensity,
$_\lambda$.
The general source function, $S_\lambda$, has been replaced by the
strict LTE
counterpart, $B_\lambda$. This means scattering has \emph{not} been included
\citep[see][for effects of scattering and a consistent formulation of it]{skartlien:bin-rad,hayek:Parallel3Dscatter}, except as absorption in the
Rosseland mean (see below).
The wavelength integration of Eq.\ (\ref{Qrad}) is approximated by the summing
over bins, $i$, reducing the problem by about four orders of magnitude compared
to a monochromatic calculation. The relative opacity,
$x_i=\kappa_i/\kappa_{\rm Ross}$, defines the bin-wise optical depth
$\tau_i = \int\varrho\kappa_{\rm Ross}x_i{\rm d}z$. For each bin, $J_i$ and
$B_i$ will diverge for $\tau_i \la 1$, giving rise to first cooling in the
photosphere, and then heating higher up in the atmosphere as $B_i$ follows the
temperature and $J_i$ converges to a constant. The
fraction of heating/cooling contribution from each bin depends largely on the
fraction of the source function, $B_i(\tau_i\sim 1)w_i$, contained in that bin.

The bin membership of each wavelength is determined from a 1D, monochromatic
radiative transfer calculation, performed on the average stratification of the
simulation, averaged over a whole number of p modes (to yield a stable
average). The bin membership is found as the Rosseland optical depth at
which the given wavelength has unity optical depth
\be{bini}
    i = {\rm int}[\log_{10}\tau_{\rm Ross}(\tau_\lambda=1)]\ ,\quad
    i \in \{0,1,2,3\}\ ,
\ee
where 'int' denotes the nearest integer. Wavelengths that go outside this
range are assigned to the continuum bin, $i=0$, or the strong-lines bin, $i=3$,
respectively. For expediency we choose to simply scale the opacity of the
continuum bin, such that $\kappa_i = 10^i\kappa_0$. The proper opacity average
within each bin, will need to converge to the diffusion approximation at
depth, and the free-streaming approximation in the high atmosphere. We therefore
choose a bridging
\be{kapbridg}
    \kappa_0 = e^{-2\tau_{\rm Ross,0}}\kappa_J 
             + (1-e^{-2\tau_{\rm Ross,0}})\kappa_{\rm Ross,0}\ ,
\ee
between the Rosseland mean of the continuum bin and mean-intensity weighted
opacity. These are defined as
\be{Ross}
    \kappa_{\rm Ross,0}^{-1} = \left.\sum_{j(i=0)}
        \frac{1}{\kappa_{\lambda_j} + \sigma_{\lambda_j}}
        \dxdy{B_{\lambda_j}}{T}w_{\lambda_j}\right/
           \!\sum_{j(i=0)}\dxdy{B_{\lambda_j}}{T}w_{\lambda_j}\ ,
\ee
where $j(i=0)$ is the set of wavelengths forming the continuum bin, according
to Eq.\ (\ref{bini}). This Rosseland mean includes both the monochromatic
absorption, $\kappa_\lambda$, and monochromatic scattering, $\sigma_\lambda$,
coefficients.

The mean-intensity weighted opacity, $\kappa_J$, is summed over all
wavelengths, but altered to suppress optically deep wavelengths with a factor
$e^{-\tau_{\lambda_j}/2}$
\be{kapJx}
    \kappa_J = \left.\sum_j\kappa_{\lambda_j}
                        J_{\lambda_j}e^{-\tau_{\lambda_j}/2}w_{\lambda_j}\right/
                 \sum_j J_{\lambda_j}e^{-\tau_{\lambda_j}/2}w_{\lambda_j}\ .
\ee
Note how scattering is \emph{not} included in $\kappa_J$, except through the
optical depth, d$\tau_\lambda=\varrho(\kappa_\lambda + \sigma_\lambda)$d$z$,
in the suppression factor.
The mean-intensity weighted opacity, $\kappa_J$, depends directly on the
1D radiative transfer solution for the mean intensity, $J_\lambda$. To cover
the whole opacity table, we extrapolate
$\kappa_J/\kappa_{\rm Ross,0}^{-1}$ from the 1D calibration
stratification to the rest of the table. This is done along fits in
$\log T$ and $\log\varrho$ to iso-optical depth contours of the simulation.
These tables are therefore specific to each individual simulation.

In plane-parallel, gray radiative transfer calculations, the atmosphere
converges to an iso-therm with height
\citep[$0.81119\stimes T_{\rm eff}$]{king:GrayAtm},
as there is no heating or cooling above $\tau\sim 0.1$. In a line
blanketed atmosphere, on the other hand, there are separate cooling peaks
from lines of a range of strengths as outlined above, resulting in atmospheres
with decreasing temperatures with height and various features arising from the
underlying opacities.

In deep layers, with continuum optical depth $\tau_1 > 300$ for all points in the plane,
we have added the radiative flux and its
associated heating, as calculated in the diffusion approximation.

\subsubsection{Radiative contributions to the EOS}

The EOS tables already include the radiative contributions in the diffusion
approximation, in particular for energy, $\varepsilon_{\rm rad}^{\rm deep}
= aT^4/\varrho$, and pressure, $p_{\rm rad}^{\rm deep} = \frac{a}{3}T^4$.
The radiation density constant is $a=8\pi^5 k^4/(15c^3h^3)$.
We use the 1D, monochromatic calibration to evaluate the proper expressions
in the atmosphere
\be{erad}
    \varepsilon_{\rm rad} = \frac{4\pi}{c}\frac{J}{\varrho}
        = \varepsilon_{\rm rad}^{\rm deep}\frac{J}{B}\ ,
\ee
and
\be{prad}
    p_{\rm rad} = \frac{4\pi}{c}K
        = p_{\rm rad}^{\rm deep}\frac{K}{B}\ .
\ee
where $K$ is the second angular moment of the specific intensity.
We therefore add
$p_{\rm rad}^{\rm deep}(\frac{K}{B}-1)$ to the pressure of the EOS table and
equivalently to the internal energy. The $J/B$- and $K/B$-ratios are
extrapolated from the 1D average to the rest of the table, as described above.

    \subsection{Relaxing the Simulations}
	\label{relax}

A simulation for a new choice of $(T_{\rm eff}, \log g)$-parameters is started from a
previous simulation with similar parameters. The physical dimensions of the
simulation box is scaled by the ratio of gravitational accelerations and the
average entropy structure is changed to result in a new $T_{\rm eff}$ based
on all the previous simulations of the grid. The behavior of the entropy in
the asymptotically deep interior, with atmospheric parameters, is shown in
Fig.~\ref{cSmax}. This asymptotic entropy is also what we feed into the
simulations through the upflows at the bottom boundary, as confirmed from
exponential fits to the horizontally averaged upflow entropy.
\mfigur{cSmax}{9.0cm}
    {The asymptotic entropy (arbitrary zero-point, see below Eq.\ [\ref{Sintegrate}]),
     $S_{\rm max}/[10^8$\,erg\,g$^{-1}$K$^{-1}$], of the deep convection zone
     as function of stellar atmospheric parameters. The $T_{\rm eff}$-scale
     is logarithmic. The entropy is indicated
     with colors as shown on the color bar, and the location of the simulations
     are shown with black asterisks, except for the solar simulation which is
     indicated with a $\odot$. For this figure only, we also added the
     simulation number from table \ref{simtab}.
     We have over-plotted tracks of stellar
     evolution computed with the MESA-code \citep{paxton:MESA}, for
     masses as indicated along each track. The dashed part shows the
     pre-main-sequence contraction, and $\alpha$ and initial helium abundance,
     $Y_0$, were determined from a calibration to the present Sun.}
The boundary affects the entropy
by prematurely pulling it up to the asymptotic value, over the bottom 4--5
grid-points (0.3--0.5 pressure scale-heights). This boundary effect on entropy
is small, though---only 0.4--1.5\% of the atmospheric entropy jump.

If we adjust only $\log g$ (with the associated scaling of the size of the box),
but keep the entropy unchanged, the new simulation will end up along the adiabat
of the original simulation and at the new $\log g$.
From Fig.~\ref{cSmax} we see that those adiabats are diagonals in the plot.
Many of the simulations
lie along such adiabats, as this is the simplest and fastest way of starting
a new simulation. The scaling of the box, should conceivably be
accompanied by some scaling of the velocities. It turns out, however, that
a factor of $10^2$ change in $g$ results in only a factor of 1.5 change in
vertical velocities (1.3 for horizontal velocities). Keeping the fluxes
consistent through the change, by not changing the velocities, seemed a better
approach. These simulations will slump or expand, necessitating a new
optimization of the vertical scale and extent.

If $T_{\rm eff}$ needs to be adjusted away from the starting simulations
adiabat, more complicated adjustments must be invoked. First we shift the
average entropy to the new $S_{\max}$ and linearly stretch the average entropy
stratification from the bottom to the atmospheric entropy minimum, to match the
entropy jump. The expected jump and $S_{\max}$ are found from
inter-/extra-polations in Figs.\ \ref{cSmax} and \ref{cSjump} between the previous simulations. We assume the
simulations to be homologous on a gas pressure scale,
$p_{\rm sc} = p_{\rm gas}/p_{\rm gas}({\rm peak~in~}p_{\rm turb})$, normalized
at the location of the maximal $p_{\rm turb}/p_{\rm tot}$-ratio. The whole
simulation cube is therefore adjusted adiabatically by the same pressure
factor, and then adjusted iso-barically to the new entropy stratification.
Our method does not rely on linearity of the
EOS, but solves numerically for entropy along pressure contours.
In both cases the changes, $\Delta\ln\varrho$ and $\Delta\varepsilon$,
are found from the average stratification only, but applied to the whole cube.

With these new pressures and densities, we scale the vertical velocities, $u_z$,to result in the projected peak $p_{\rm turb}/p_{\rm tot}$-ratio. We then
adjust the amplitude of the internal energy fluctuations (keeping all the
carefully adjusted averages unchanged) in order to reproduce the target
convective flux. We find a hydrostatic $z$-scale by inverting the equation of
hydrostatic equilibrium
\be{hydrostatz}
    \dd{P}{z} = g\varrho\quad\Leftrightarrow\quad
        z =\int_{P_{\rm tot,\,bot}}^{P_{\rm tot}(z)}
                \frac{{\rm d}P_{\rm tot}}{g\varrho}\ ,
\ee
and integrating from the bottom and up. This $z$-scale will be rugged and not
optimal for resolving the hydro- and thermo-dynamics. The last step is therefore
to compute an optimized $z$-scale and interpolate the simulation cubes to this.
This procedure results in simulations that are rather close to their
(quasi-static) equilibrium state, minimizing the relaxation time.

No matter how carefully such a scaled simulation has been constructed, it
will still not be in its new (quasi-static) equilibrium
configuration, since we do not
know {\it a priori} what that equilibrium is---that is the whole reason we need to
perform these simulations, after all. The natural state is the lowest energy
state, so departure from equilibrium means an excess of energy. This extra
energy is quickly spent on exciting p modes (sound waves) and over a few sound
crossing times, the energy gets channeled into 3--4 of the lowest order
modes of the simulation box (radial, as well as non-radial). We extract
surplus energy from the simulation by
damping the radial p modes by adding a term
\be{pdamp}
    -\frac{v_{\rm mode}}{t_{\rm damp}}\ , {\rm with}\quad
    v_{\rm mode} = \frac{\langle\varrho u_z\rangle}{\langle\varrho\rangle}\ ,
\ee
to the radial part of the momentum equation, Eq.~(\ref{NavStok}b).
For the damping time-scale, $t_{\rm damp}$, we use 1.3 times the period of the
fundamental p mode, $P_1$. When the center-of-mass velocity of the
simulation no longer displays simple sinusoidal modes, the damping time is
increased smoothly (exponentially) until the damping is insignificant at which
point it is turned off. The simulation is run without damping for another $2P_1$
for which we compute horizontal averages of $\varrho$ and $T$.
This constitutes the first iteration of the relaxation process.

If the stratification has changed from the previous iteration, the opacity
binning is re-calculated with the new stratification and we start a new
iteration of the relaxation process (from Eq.~(\ref{pdamp}). Otherwise we can
start the production runs, from which the results in this paper are based.
Each production run covers at least $10P_1$ and all snapshots (about 100--300)
of each simulation are included in the averaging producing our results.

Some further requirements for a relaxed simulation are that the fluxes have
reached a statistically steady state, that the total flux is statistically
constant, that no quantities (especially at the bottom boundary) show any
significant drift, that the radiative heating is statistically constant and
that the time-steps are like-wise. All these precautions ensure that we have
a homogeneous set of simulations and that we can trust in the differences
between the simulations and therefore can trust the dependencies of derived
quantities with $T_{\rm eff}$ and $\log g$.
    \subsection{The Grid of Simulations}
	\label{thegrid}

Our 37 simulations span most of the convective range of the zero age main
sequence (ZAMS) from $T_{\rm eff}=4$\,300\,K to 6\,900\,K (K5--F3) and up to
giants of $\log g = 2.2$ between $T_{\rm eff}=5$\,000\,K and 4\,200\,K (K2--K5),
\begin{table*}[htbp]\scriptsize
  \caption{\label{simtab} Fundamental parameters for the 37 simulations, and
  a few derived quantities.\tablenotemark{*}}
  \begin{flushleft}\begin{tabular}{rcrcrrrrrrrrr}
            \noalign{\vskip -4.0ex}
      \hline\noalign{\vskip 0.5ex}
      \hline\noalign{\smallskip}
  \oh{sim\tablenotemark{a}} & {MK\,class\tablenotemark{b}\hspace{-1.5em}}
& \oh{$\frac{\displaystyle T_{\rm eff}}{\displaystyle\rm[K]}$}
& \oh{$\log g$}
& \oh{$\frac{\displaystyle w\tablenotemark{c}}{\displaystyle\rm[Mm]}$}
& \oh{$\frac{\displaystyle d\,\tablenotemark{d}}{\displaystyle\rm[Mm]}$}
& \oh{$\frac{\displaystyle\Delta z\tablenotemark{e}}{\displaystyle\rm[Mm]}$}
& \oh{$\frac{\displaystyle A_{\rm gran}\tablenotemark{f}}{\displaystyle\rm[Mm]}$}
& \oh{$\frac{\displaystyle\Lambda\tablenotemark{g}}{\displaystyle\rm[Mm]}$}
& \oh{$S_{\rm max}$\tablenotemark{h}}
& \oh{$S_{\rm jump}$\tablenotemark{h}}
& \oh{$\frac{\displaystyle P_{\rm turb}}{\displaystyle P_{\rm tot}}$}
& \oh{$\frac{\displaystyle I_{\rm RMS}}{\displaystyle\langle I\rangle}$} \\
      \noalign{\smallskip} \hline \noalign{\smallskip}
  1 &     K3 & $  4681\pm 19$ & 2.200 & 1248.30 & 652.410 &  3.478--16.695 & 283.010 &   78.495 &  7.9539 &  6.7934 & 0.26223 & 0.18555 \\
  2 &     K2 & $  4962\pm 21$ & 2.200 & 1800.00 & 958.370 &  4.598--27.616 & 382.270 &  170.020 & 11.3540 &  9.5480 & 0.29639 & 0.18758 \\
  3 &     K5 & $  4301\pm 17$ & 2.420 &  667.81 & 409.900 &  1.724--12.100 & 136.080 &   32.745 &  3.9781 &  3.2772 & 0.19130 & 0.17758 \\
  4 &     K6 & $  4250\pm 11$ & 3.000 &  165.00 &  78.271 &  0.457-- 1.815 &  31.953 &    3.805 &  1.6903 &  1.6152 & 0.14336 & 0.16793 \\
  5 &     K3 & $  4665\pm 16$ & 3.000 &  169.47 & 115.880 &  0.649-- 2.564 &  36.060 &    6.236 &  3.1968 &  2.7581 & 0.16881 & 0.17371 \\
  6 &     K1 & $  4994\pm 15$ & 2.930 &  275.67 & 132.840 &  0.591-- 3.752 &  50.140 &    9.476 &  5.1703 &  4.4027 & 0.20124 & 0.17806 \\
  7 &     G8 & $  5552\pm 17$ & 3.000 &  328.29 & 160.950 &  0.787-- 4.101 &  52.785 &   17.077 &  8.7944 &  7.5617 & 0.26233 & 0.19568 \\
  8 &     K3 & $  4718\pm 15$ & 3.500 &   52.18 &  30.682 &  0.176-- 0.751 &  11.005 &    1.203 &  1.7158 &  1.7188 & 0.13479 & 0.15879 \\
  9 &     K0 & $  5187\pm 17$ & 3.500 &   53.59 &  36.937 &  0.196-- 0.891 &  11.372 &    1.826 &  3.1948 &  2.8700 & 0.17238 & 0.17183 \\
 10 &     K0 & $  5288\pm 20$ & 3.421 &   89.00 &  38.936 &  0.205-- 0.958 &  15.609 &    2.391 &  3.9879 &  3.5133 & 0.17966 & 0.17845 \\
 11 &     F9 & $  6105\pm 25$ & 3.500 &   71.01 &  52.645 &  0.220-- 1.416 &  17.123 &    5.964 &  9.2519 &  8.0430 & 0.27125 & 0.20493 \\
 12 &     K6 & $  4205\pm\p8$ & 4.000 &   14.83 &   6.342 &  0.039-- 0.138 &   2.753 &    0.116 & -0.3701 &  0.5958 & 0.07684 & 0.11957 \\
 13 &     K4 & $  4494\pm\p9$ & 4.000 &   14.83 &   6.420 &  0.046-- 0.113 &   2.981 &    0.165 &  0.1071 &  0.7660 & 0.09391 & 0.13550 \\
 14 &     K3 & $  4674\pm\p8$ & 4.000 &   14.83 &   7.437 &  0.034-- 0.204 &   2.945 &    0.216 &  0.4342 &  0.9427 & 0.10296 & 0.13596 \\
 15 &     K2 & $  4986\pm 13$ & 4.000 &   16.50 &   9.026 &  0.058-- 0.196 &   3.273 &    0.299 &  1.0546 &  1.3418 & 0.11742 & 0.14480 \\
 16 &     G6 & $  5674\pm 16$ & 3.943 &   19.33 &  12.785 &  0.069-- 0.285 &   4.069 &    0.663 &  3.1889 &  2.9713 & 0.17173 & 0.17943 \\
 17 &     F9 & $  6137\pm 14$ & 4.040 &   21.40 &   9.326 &  0.041-- 0.300 &   4.019 &    0.747 &  4.9301 &  4.4363 & 0.20478 & 0.20117 \\
 18 &     F4 & $  6582\pm 26$ & 3.966 &   24.28 &  18.246 &  0.092-- 0.421 &   6.034 &    2.087 &  9.2444 &  8.1239 & 0.26844 & 0.21381 \\
 19 &     F4 & $  6617\pm 33$ & 4.000 &   22.45 &  16.877 &  0.085-- 0.389 &   5.368 &    1.936 &  9.2407 &  8.1188 & 0.25049 & 0.21202 \\
 20 &     K4 & $  4604\pm\p8$ & 4.300 &    7.43 &   3.625 &  0.017-- 0.101 &   1.405 &    0.076 & -0.2001 &  0.6473 & 0.08236 & 0.12048 \\
 21 &     K1 & $  4996\pm 17$ & 4.300 &    7.43 &   3.772 &  0.019-- 0.096 &   1.515 &    0.113 &  0.4301 &  0.9887 & 0.10066 & 0.13029 \\
 22 &     K1 & $  5069\pm 11$ & 4.300 &    8.27 &   3.531 &  0.023-- 0.074 &   1.615 &    0.115 &  0.5441 &  1.0562 & 0.10300 & 0.13118 \\
 23 &     K0 & $  5323\pm 16$ & 4.300 &    8.27 &   4.135 &  0.023-- 0.096 &   1.614 &    0.145 &  1.0413 &  1.3973 & 0.11662 & 0.14201 \\
 24 &     G1 & $  5926\pm 18$ & 4.295 &    9.32 &   5.473 &  0.024-- 0.139 &   1.872 &    0.259 &  2.6796 &  2.6358 & 0.15869 & 0.18234 \\
 25 &     F5 & $  6418\pm 26$ & 4.300 &   11.76 &   5.373 &  0.030-- 0.128 &   2.221 &    0.406 &  4.9238 &  4.4645 & 0.20351 & 0.20745 \\
 26 &     F2 & $  6901\pm 29$ & 4.292 &   11.45 &   8.334 &  0.037-- 0.232 &   2.760 &    0.978 &  9.2408 &  8.1679 & 0.27642 & 0.20984 \\
 27 &     K4 & $  4500\pm\p4$ & 4.500 &    4.69 &   2.108 &  0.013-- 0.051 &   0.886 &    0.031 & -0.6094 &  0.4978 & 0.06359 & 0.11046 \\
 28 &     K3 & $  4813\pm\p8$ & 4.500 &    4.69 &   2.026 &  0.010-- 0.054 &   0.949 &    0.049 & -0.2102 &  0.6576 & 0.07991 & 0.11609 \\
 29 &     K0 & $  5232\pm 12$ & 4.500 &    4.69 &   2.344 &  0.011-- 0.062 &   1.017 &    0.071 &  0.4277 &  1.0217 & 0.10218 & 0.12981 \\
 30 &     G5 & $  5774\pm 17$ & 4.438 &    6.02 &   3.476 &  0.020-- 0.082 &   1.237 &    0.140 &  1.7112 &  1.9149 & 0.13747 & 0.16345 \\
 31 &     F7 & $  6287\pm 15$ & 4.500 &    5.36 &   3.498 &  0.020-- 0.074 &   1.178 &    0.179 &  3.1827 &  3.0738 & 0.16893 & 0.19410 \\
 32 &     F4 & $  6569\pm 17$ & 4.450 &    8.33 &   3.704 &  0.018-- 0.097 &   1.538 &    0.288 &  4.9211 &  4.4866 & 0.20431 & 0.20658 \\
 33 &     K1 & $  5021\pm 11$ & 4.550 &    4.18 &   2.055 &  0.012-- 0.050 &   0.850 &    0.052 &  0.0039 &  0.7731 & 0.08862 & 0.11827 \\
 34 &     G9 & $  5485\pm 14$ & 4.557 &    4.17 &   2.244 &  0.013-- 0.050 &   0.849 &    0.073 &  0.7452 &  1.2347 & 0.11311 & 0.13862 \\
 35 &     G1 & $  5905\pm 15$ & 4.550 &    4.65 &   2.547 &  0.015-- 0.063 &   0.933 &    0.105 &  1.7012 &  1.9247 & 0.13971 & 0.16830 \\
 36 &     K6 & $  4185\pm\p3$ & 4.740 &    2.70 &   1.016 &  0.006-- 0.025 &   0.488 &    0.009 & -1.6403 &  0.3562 & 0.03898 & 0.07347 \\
 37 &     K4 & $  4531\pm 10$ & 4.740 &    2.70 &   1.140 &  0.007-- 0.025 &   0.539 &    0.014 & -0.8481 &  0.4077 & 0.05513 & 0.09950 \\
    \noalign{\smallskip} \hline
    \noalign{\vskip -2.5ex}
  \end{tabular}
\setlength{\columnwidth}{513pt}
\tablenotetext{*}{This table is published in its entirety in the
electronic edition of the {\it Astrophysical Journal}.}
\tablenotetext{a}{The simulation numbers are also shown in Fig.\ \ref{cSmax}.}
\tablenotetext{b}{The spectral class is only approximate, as it depends on the
    luminosity.}
\tablenotetext{c}{Width of the simulation domain, $w$.}
\tablenotetext{d}{Depth of the simulation domain, $d$.}
\tablenotetext{e}{Min.\ and max.\ vertical grid-spacing.}
\tablenotetext{f}{The typical granulation size, $A_{\rm gran}$ (See Sect.\
    \ref{varTg}).}
\tablenotetext{g}{The convective expansion of the atmosphere, $\Lambda$ (See
    Sect.\ \ref{SurfEff}).}
\tablenotetext{h}{The unit of entropy is $10^8$\,erg\,g$^{-1}$\,K$^{-1}$, and
    the zero-point is specified below Eq.\ (\ref{Sintegrate}).\vskip 0.0ex}
\setlength{\columnwidth}{245.26653pt}
  \end{flushleft}
\end{table*}
%
as listed in table \ref{simtab}. The location of the simulations in the
atmospheric HR-diagram is also shown in Fig.\ \ref{cSmax}.

Our grid of simulations is not regular, for a couple of
practical reasons. The most important one being that $T_{\rm eff}$ is not an
actual parameter of the simulations. We only have indirect control over
$T_{\rm eff}$, by adjusting the entropy of the upflows entering through the
bottom boundary.
Requiring particular values of $T_{\rm eff}$ would add another outer loop of
iterations of the relaxation process (Sect.~\ref{relax}), for monitoring and
adjusting the resulting $T_{\rm eff}$. Since this takes extra (human) time we
have not required a regular grid and pre-determined values of $T_{\rm eff}$.
This is the reason that many of the simulations lie along adiabats of the deep
entropy, as shown in Fig.~\ref{cSmax}. Those simulations result from changes to
$\log g$ (with the associated scaling of the size of the box) only, without
adjusting the entropy structure. The entropy structure, of course, changes
during the subsequent relaxation of the simulation, but the asymptotic deep
value is fixed.

Not insisting on a regular grid, also has the advantage of being able to
have higher resolution around the main sequence, where it is most needed.
This makes it possible to resolve features
that would be missed in a coarser, regular grid.

\subsubsection{Interpolating in the Grid}

Interpolation in such an irregular $(\log T, \log g)$-grid is a bit more complicated compared to
other grids of stellar atmospheres.
Fortunately the triangulation of arbitrarily distributed
points in a plane is not a new discipline. We use the suite of Fortran77
routines developed by \citet{renka:triangulation}, which performs both the
triangulation and subsequent interpolations by piece-wise cubic 2D functions.
The routines can also return the partial derivatives of the interpolation.
The triangulation routines supplied in
IDL\footnote{IDL{\raise0.9ex\hbox{\tiny\textregistered}} is a registered
trademark of ITT Visual Information Solutions, USA.} are also based on
\citet{renka:triangulation}, facilitating visualization and interactive
debugging.

It makes no sense to interpolate individual snapshot of whole simulation cubes
between different simulations, since they will not be homologous. We therefore
only interpolate between temporally and horizontally averaged structures, or
between derived quantities, {\eg}, limb darkening coefficients or the intensity
contrast of granulation.
	\subsubsection{Averaging Procedures}
	\label{av-proc}

The averaging of the simulations is carried out in two ways.
First, a horizontal averaging mapped onto a (horizontally averaged) column
density scale in order
to filter out the main effect of the radial p-modes excited in the simulations.
We call this pseudo Lagrangian averaging and denote it by
$\langle\dots\rangle_{\rm L}$. The reason for the ``pseudo'' is the use of a
1D average column density scale, as opposed to the local column density.

The second method we employ, is an optical depth averaging performed by
averaging the quantities in
question, over the undulating iso-$\tau$ surfaces. This average we denote by
$\langle\dots\rangle_\tau$ and it has been calculated for both the Rosseland
optical depth, $\tau_{\rm Ross}$ and the 5\,000\,{\AA} monochromatic optical
depth, $\tau_{5000}$.

What kind of average, to use, depends on the problem at hand, and should be
guided by the structure of the equations involved (before approximations and
simplifications are made).

	\section{General Properties of Surface Convection}
	\label{GenConv}

Deep convection is very nearly adiabatic (to parts per million), whether it
occurs in the core of stars, or in the envelope. Here, no theory of convection
is necessary for determining the structure of the star, as the stratification just follows
an adiabat. This property makes deep convection very attractive for 
EOS research \citep{dappen:SunSeismHe,dappen:SunEOSlab}, since the adiabat is
completely determined by the EOS, and the matter is fully mixed, greatly
simplifying the analysis.

At the top of convective envelopes, the situation is 
much more complicated, not the least
because it overlaps with the photospheric transition from optical deep
(diffusion approximation) to optical thin (free-streaming approximation) and
is accompanied by an abrupt change in density gradient. We outline the details
below \citep[also see][]{bob:conv-topology,aake:santa-barbara,nordlund-stein:score96,nordlund:LivingRev}.

The upflows are close to isentropic, rising from the deep interior where
convection is adiabatic.
They also exhibit only little turbulence, since they are expanding along the
exponential density gradient, smoothing out any fluctuations. The smoothing
effect gets stronger as the density gradient and velocities increase towards
the surface. This density gradient
also forces the upflow to continuously overturn into the downdrafts, in order
to conserve mass under the constraint of hydrostatic equilibrium. This
overturning occurs with a scale-height of a mass mixing length,
\be{eq:alpham}
    \ell_m = \left|{\rm d} \ln F_{\uparrow m}(r) / {\rm d}r \right|^{-1}\ ,
\ee
where $F_{\uparrow m}(r)$ is the mass flux, $\varrho u_z$, averaged over the
upflows only \citep[][TS11]{trampeda:mixlength}. This definition means a
fraction of $e^{-1}$ of the upflow will have overturned into the downdrafts
over the height interval, $\ell_m$. In the deeper parts of our simulations
(the same that were analyzed by TS11), below the superadiabatic peak, the mass
mixing length is proportional to the pressure scale height, $\ell_m = \alpha_m
H_p$ (not the density scale-height, $H_\varrho$, nor the distance to the top of
the convection zone, $\Lambda$, as has been proposed in the past).
We show this in Fig.\ \ref{alpham} for three of our simulations that span
\mfigur{alpham}{6.5cm}
    {The mass mixing-length, $\ell_m$, in units of pressure scale-height, $H_p$,
     for three of our simulations (See table \ref{simtab}), as presented by
     \citet{trampeda:mixlength}. The abscissa is the logarithmic total pressure
     normalized by the pressure, $p_{\rm eff}$, at the depth where
     $\langle T\rangle=T_{\rm eff}$. The dashed, horizontal lines show the range
     of $\ell_m/H_p$ determined for each of the three simulations. The
     vertical dotted lines show the location of the top of the respective
     convection zones. The four asterisks on each curve, show the position of
     the temperature snapshots shown in Fig.\ \ref{gran_form}.}
our grid. TS11 finds
values of the deep $\alpha_m$ between 1.67 and 2.20.
The sharp rise in  $\alpha_m$ at the bottom, is due to boundary effects.
The mass flux is the quantity displaying the most extreme boundary effects,
being further amplified by the differentiation, in this case.

This simple, but powerful concept, of convective flows being dominated by
continuous overturning of adiabatic upflows due to mass conservation of flows
along a density gradient, is also predictive. At depths where radiative losses
are negligible, pure advection of the entropy deficit from surface cooling into
an isentropic interior, would result in a local entropy deficit,
$S(z)-S_{\max}$, that scales with the density ratio, $\varrho(z=0)/\varrho(z)$.
This is confirmed to high accuracy by the simulations (below $z=1$\,Mm for the
solar case).

There is a peak in $\ell_m/H_p$ about a pressure-decade below the top of the
convection zone, before it drops off to about 1 above the photosphere. The top
of the convection zone is evaluated as the depth at which the convective flux
($F_{\rm conv} = F_H + F_{\rm kin}$, the sum of enthalpy and kinetic fluxes) is
zero, and is indicated with the vertical dotted lines in Fig.\ \ref{alpham}.
This point is well-defined since $F_{\rm conv}$ is
negative in the overshoot region above the convection zone.
For comparison, the superadiabatic gradient peaks around $p/p_{\rm eff}\sim
1$, straddling the photosphere.

The sharp drop in $\ell_m$ from its peak value, reflects a
change in morphology of the flows. As the mixing-length decreases faster
than the pressure scale-height the same upflows have to overturn over a
decreasing height-scale, forcing
the whole circumference, and not just the vertices, of the upflows to be
efficient downflows --- this is
what we recognize as granulation in the solar photosphere. This granulation
pattern is formed in the front side of the mass mixing-length peak, as
\mfigur{gran_form}{12.7cm}
    {Horizontal planes of temperature for arbitrary snapshots of the three
     simulations shown in Fig.\ \ref{alpham}, and at the heights indicated
     with asterisks in that figure. The top row shows disk-center, white-light
     intensity. For each row the images are scaled so that contrasts can be
     directly compared, but the magnitude between simulations can not.}
shown in Fig.\ \ref{gran_form}. Here we show the change of convective
morphology with depth, for simulation 37 (sedate convection in a cool MS
dwarf), 26 (vigorous convection in a warm MS star) and 2 (vigorous convection
in a red giant). The top row shows the emergent granulation pattern in
disk-center, white-light intensity.
The third and fourth rows show approximately the top and bottom of
the granules and the second and bottom rows show convection six grid-points above
and below that, respectively, to illustrate the transition away from granular
convection. The displayed quantity is temperature, but it is very similar
in internal energy. Vertical velocity, or velocity divergence turned out to be less
well suited for finding granules. We (loosely) define the bottom of the
granules as where the cooler inter-granular lanes start to disconnect and and
the downflows concentrate into knots connected by weaker lanes.

In the layer at the top of the granulation (third row) we see all of the
granules in the most sedately convective simulation \# 37. The most vigorous
convection, as seen in \# 26, is accompanied by a sharper photospheric
temperature drop and we only see a few granules at the
peak temperature (brightest in Fig.\ \ref{gran_form}, third row) with the rest
of the granules simply peaking in slightly deeper layers. The second row of
images shows the morphology at the top of the convection zone, which is
characterized by hot lanes
in the middle of inter-granular lanes, and some granules being cooler
than these hot lanes at this height. These hot lanes are optically thin,
however, and contribute little to the emergent picture of granulation, while
the contribution by the granules is just fading, as only the hottest parts
of the strongest granules have their photospheres at this height.

The above described formation of granules is a very
generic result. Whenever convection takes place in a stratified medium and the
mass mixing length undergoes a sudden drop, faster than the pressure
scale-height
(typically by some form of cooling) a granulation pattern will emerge.
It occurs in simulations with
simple gray radiative transfer \citep{kim:shallow-conv,robinson:3DSun}, in simulations with a
completely artificial cooling layer and with impenetrable upper boundary
\citep[as in the interior convection simulations of, {\eg},][]{miesch:SunASH,brun:SolarAsh},
and it occurs in laboratory experiments with Rayleigh-B{\'e}rnard convection
\citep{koschmieder:LabRaylBenConv,meyer:LabRaylBenConv,busse:LabRaylBenConv},
where the mixing length drops to zero at the surface of the fluid.

In real stellar convection, however, the contrast between the upflows
and the downdrafts is controlled by the details of the radiative cooling
in the photosphere.
This is dominated by the highly temperature sensitive H$^{-}$ opacity ($\propto
T^9$). The isentropic upflows therefore only start cooling
very close (spatially) to the unity optical depth. In the cooler downdrafts, on the
other hand, unity optical depth occurs much deeper (on a spatial scale)
\citep{georgobiani:p-mode-assym}
and the cooling also occurs over a larger height range. The cooled downflows
are also continuously
mixed with the high-entropy plasma overturning from the upflows. As they move
against the density gradient, the downflows are also turbulent and they
contain the majority of vorticity---horizontal at the periphery of downdrafts and
vertical in the centers. This inherent
asymmetry between the upflows and the downdrafts makes it rather difficult
to treat in analytical 1D formulations. The asymmetry also results in density
differences between upflows and downdrafts, which gives rise to a kinetic
energy flux which is negative and increasing to about 20\% of the total flux
at the bottom of our simulation domains. This kinetic flux is not accounted
for in MLT formulations, but together with the enthalpy flux makes up the
convective flux in the simulations. There are also 5--13\% contributions from
acoustic fluxes inside the convection zone (largest for cool giants) and 
minor viscous fluxes that peak at less than 1\%, just below the superadiabatic peak.
The density difference in turn implies
a difference in filling factor between the up- and downflows (assuming
hydro\-static equilibrium as realized to a large extent by the simulations).
The filling factor is about 60\% of the area being in the upflows in the deep
parts, fairly constantly with height and atmospheric parameters. It changes
abruptly at the top of the convection zone, to be a bit below 50\% above the
photosphere.

\subsection{Convective Efficiency}
\label{ConvEff}

The efficiency of convection is intimately connected to the convective
velocities and the temperature and density contrasts. A larger atmospheric
entropy jump (a larger superadiabatic peak) means less efficient convection,
as the deviation from an adiabat implies energy losses on the way. The ideally
efficient convection would not lose any energy before reaching the top of the
convection zone. Comparing two cases with differing efficiency,
but same total flux, the less efficient will necessarily have higher
convective velocities in order to transport the same flux. These larger
vertical velocities also means larger horizontal, overturning velocities
between the upflows and the downdrafts, sustaining larger pressure contrast
between the two. The entropy jump is set up by increased cooling around the
photosphere, which drives a larger temperature contrast between the upflows
and the downdrafts (and therefore also a larger emergent intensity contrast
between granules and inter-granular lanes, as seen in Fig.~\ref{cIRMS}).
This entropy jump is shown in Fig.\ \ref{cSjump} for our grid of simulations
\mfigur{cSjump}{9.0cm}
    {As Fig.\ \ref{cSmax}, but showing the atmospheric entropy jump
     $S_{\rm jump}/[10^8$\,erg\,g$^{-1}$K$^{-1}]$,
     a measure of convective efficiency, as function of
     stellar atmospheric parameters.}
\citep[also see][]{trampedach:Innsbruck2012}. We clearly see that cool dwarfs
have the most efficient convection and it gets progressively less efficient
as the limit of convective envelopes is approached (a diagonal in the plot,
above the high-$T_{\rm eff}$/low-$g$ limit of our grid). Beyond this limit no
significant flux is transported by convection near the surface.
Comparing with the asymptotic adiabat shown in Fig.\ \ref{cSmax} (or the
turbulent pressure ratio in Fig.\ \ref{cPtmax}), we see a very
similar behavior, although the entropy jump falls off faster as the limit of
convective envelopes is approached.  In terms of MLT formulations, a large
convective efficiency is accomplished with a large mixing length, which carries
entropy further before mixing with the surroundings.

Two stars that are otherwise similar (in effect, the same $\log g$ and
asymptotic entropy, $S_{\max}$), but have different atmospheric opacities,
will also have different convective efficiencies. The radiative loss of energy
around the photosphere, will happen in the same range of optical depth, for
the two stars, but for the star with lower opacity this range will correspond
to a larger range in (spatial) depth, as ${\rm d}\tau=\varrho\kappa{\rm d}z$.
With similar convective velocities, this
would lead to more time spent loosing energy to the radiation field, and
consequently increasing the atmospheric entropy jump and hence, decreasing the
convective efficiency. A lower efficiency leads to higher velocities, which
will limit the radiative losses above, providing a negative feed-back and
enabling an equilibrium state. The net flux and therefore $T_{\rm eff}$ will
most likely be affected by such an opacity change.
One reason for lower atmospheric opacities, is of course
a lower metallicity, which would then be expected to result in lower convective
efficiency. We do not explore variation in metallicity with our grid of
simulations, but a recent investigation of solar like Kepler stars by
\citet{bonaca:KeplerAlpha}, finds a decreasing $\alpha$ with decreasing
metallicity, as argued above. This opacity effect on convection is, of course,
also relevant for assessing effects of opacity updates on models or of how
composition altering processes could affect stars and maybe explain
observations.

\subsection{Seismology and Surface Convection}
\label{SurfEff}
%

The so-called surface effect \citep{jcd:solar-freq-shifts}, is a systematic and
persistent difference between observed and model frequencies, which is
independent of the degree, $l$, of the modes.
This surface effect is caused by differences between the star and the model
in the layers around the upper turning-point of the p modes. This is also the
layers where the photospheric transition from optical deep (diffusion
approximation) to optical thin (free streaming), and the transition at the
top of the convection zone from convective to radiative transfer of the flux,
occur.
The latter results in the most inefficient convection in that star, with large
velocities, large turbulent pressure, large temperature fluctuations and the
most superadiabatic convection. The photospheric transition enhances the
convective fluctuations by rapid cooling of the upflows. All these effects,
and the strong coupling between radiation and convection, means that a 1D,
plane-parallel, MLT atmosphere is an inadequate description. Most of the
assumptions that MLT build on are violated in this region and we therefore
expect, and have seen, large systematic differences between observed and model
p mode frequencies. Since the problems occur in a thin layer around the upper
turning point, all modes feel it equally, and the frequency differences are
therefore mainly a function of frequency. The usual procedure
is therefore to adjust the model frequencies by a smoothly varying function
of frequency. This procedure seems to work well for then Sun, and has also
been applied successfully to other stars \citet{hans:SeismNearSurf}.

The main reason for the surface effect is a convective expansion of the
\mfigur{cExpan}{9.0cm}
    {As Fig.\ \ref{cSmax}, but showing the logarithmic total convective
     expansion of the atmosphere, $\Lambda_{\rm conv}$, as function of stellar
     atmospheric parameters.}
atmosphere, which enlarges the acoustic cavity of the modes. This expansion
is shown in Fig.\ \ref{cExpan} for the grid of simulations. It is evaluated
as the average outward shift of the (total) pressure stratification between
$\langle\ln p_{\rm tot}\rangle_{\rm L}$ of each simulation and their
corresponding 1D stellar envelope model, from their respective photospheres
and up. The 1D models are based on the same EOS and opacities, and have been
matched to the averaged simulations, thereby calibrating the mixing-length
parameter, $\alpha$ \citep{trampedach:alfa-fit}. This ensures the 3D and 1D
models are comparable, with the only major difference being the treatment of
convection.

About half of this expansion is due to the turbulent pressure contributing to
the hydrostatic support. The maximum of the turbulent- to total-pressure ratio is shown in
Fig.\ \ref{cPtmax}, and is found to depend on atmospheric parameters in a
similar way as the entropy of the asymptotic adiabat in Fig.\ \ref{cSmax}.
Fig.\ \ref{cPtmax} spans pressure ratios from 3.90\% for the coolest dwarf
to 29.6\% for the warmest giant (closely followed by 27.6\% for the hottest
dwarf in our grid.
\mfigur{cPtmax}{9.0cm}
    {As Fig.\ \ref{cSmax}, but showing the maximum (with depth) of the
     $P_{\rm turb}/(P_{\rm turb}+P_{\rm gas})$-ratio, as function
     of stellar atmospheric parameters.}
From the different $T_{\rm eff}$ and $\log g$ dependencies in
Figs.\ \ref{cExpan} and \ref{cPtmax} it is clear that it is not the pressure
ratio itself that determines the atmospheric expansion.
As we did for Eq.\ (\ref{hydrostatz}), we invert the equation of hydrostatic equilibrium and now integrate over just the turbulent component of the pressure,
to get the associated expansion
\be{LmbdTurb}
    \Lambda_{\rm turb}(z) =\int_{P_{\rm turb}(z_{\rm bot})}^{P_{\rm turb}(z)}
            \frac{{\rm d}P_{\rm turb}(z^\prime)}{g\varrho(z^\prime)}\ .
\ee
The total pressure is composed of, e.g., $P=P_{\rm rad} + P_{\rm gas}
+ P_{\rm turb}$, and the gas pressure can be further decomposed into
contributions by individual elements, ions, electrons or molecules. The
amount of atmospheric expansion by any of those pressure components can be computed 
in the same way as that by the turbulent pressure, in Eq.\ (\ref{LmbdTurb})
above. The
integration is carried out from the interior to the top of the domain. About
half of $\Lambda_{\rm turb}$ ($\sim$quarter of the full $\Lambda_{\rm conv}$) is
actually supplied by the overshoot region above the convection zone, which in
most 1D MLT models has no velocity fields.

This turbulent component provides approximately half of the full expansion
shown in Fig.\ \ref{cExpan}. The rest is due to the so-called \emph{convective 
back-warming} \citep[Paper\,I]{trampedach:SOHO12}, which is caused by the large
convective temperature fluctuations in the photosphere, combined with the
high temperature sensitivity of the opacity there ($\kappa({\rm H}^-)\propto
T^9$). This effectively limits the radiative losses from the hot granules, to
a larger extent than would have been the case for a corresponding 1D model with
the average temperature stratification. Because the temperature fluctuations are well past
the linear regime of the opacities, the effect is not symmetrically opposite
in the cooler inter-granular lanes, and a net back-warming results, akin to the
effect of line-blanketing \citep[Sect.\ 6.1]{MARCS-2}. The back-warming
results in a larger (gas) pressure scale-height, {\ie}, an expansion of the
atmosphere.

\begin{figure}[hbt]\leftline{\psfig{figure=\figdir/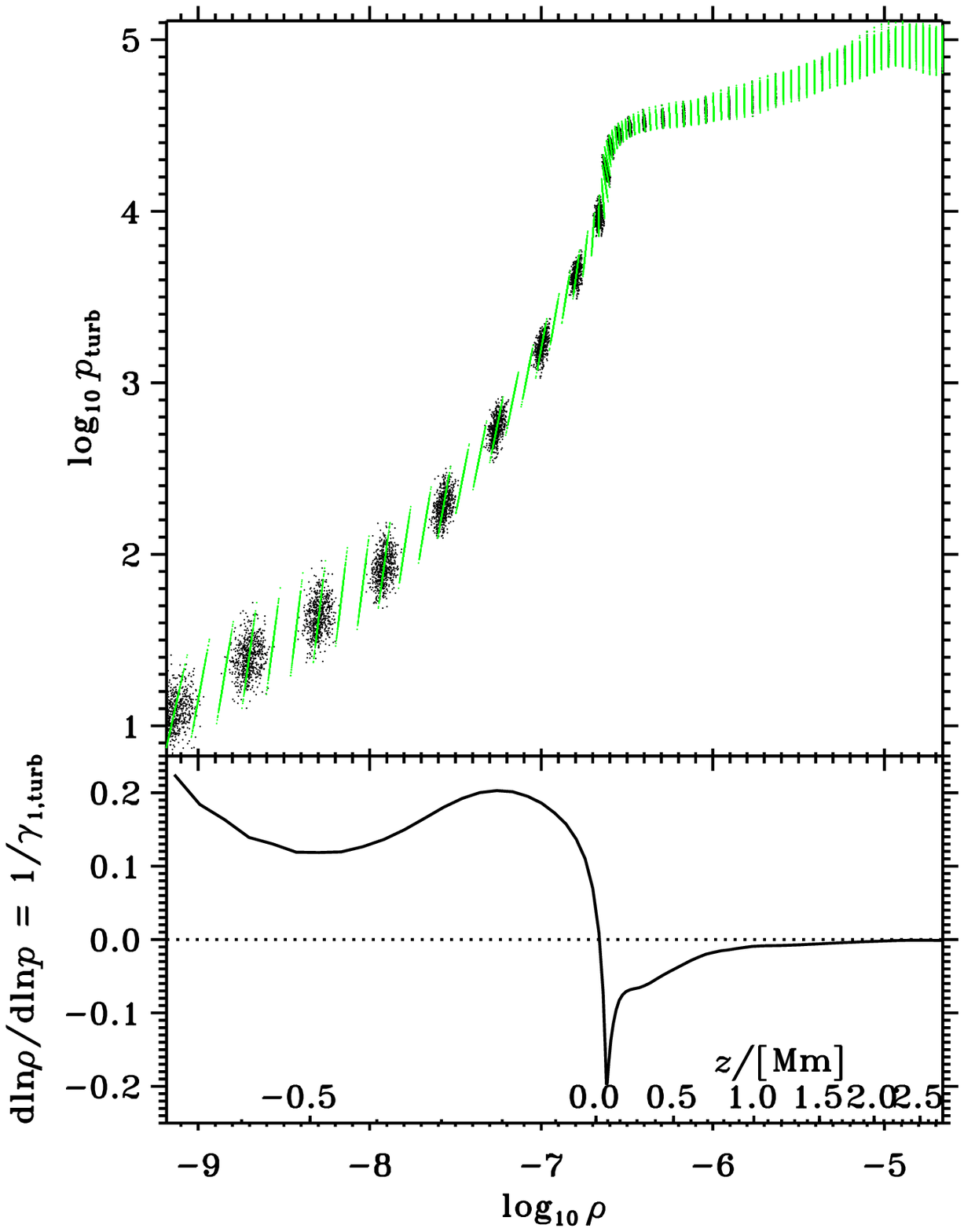,height=10cm}}
    \caption{\label{gm1turb} {\bf Top panel:} Black points show horizontally
     averaged $\langle p_{\rm turb}(t)\rangle_{\rm L}$ vs.
     $\langle\varrho(t)\rangle_{\rm L}$ for each
     snapshot of the solar simulation time-series (for every third depth-point).
     The green lines show linear fits to these points, at each depth.
     {\bf Bottom panel:} Inverse gradients of the fits from the top panel, as
     function of depth (upper axis) and density (lower axis, in common with the
     top panel.}
\end{figure}

\citet{rosenthal:conv-osc} found 
the expansion of the solar atmosphere, relative to a 1D MLT model, to be a
major ingredient of the surface effect. The effective gradient of
the turbulent pressure, $\gamma_{1,{\rm turb}}$ \citep{bob:santa-barbara},
however, was found to be of similar importance. We show the inverse of this,
${\rm d}\ln\varrho/{\rm d}\ln p_{\rm turb}$ for the solar simulation in
Fig.\ \ref{gm1turb}. From the top panel of this plot, we see how
$\langle p_{\rm turb}(t)\rangle_{\rm L}$ vs. $\langle\varrho(t)\rangle_{\rm L}$
displays a great deal of scatter, increasing with height above the photosphere.
Despite this, a linear fit at each height is well-defined and gives a rather
steep positive gradient above the photosphere, going through vertical in the
photosphere (the reason we show $1/\gamma_{1,{\rm turb}}$ in the bottom panel),
and being steep and negative in the interior, converging towards vertical (with
minimal scatter). The shape of $1/\gamma_{1,{\rm turb}}$ is rather universal
for our simulations, mainly varying in the amplitude of the various components
of the curve.

The appropriate form of the total $\gamma_1$ can be gleaned from the manner
in which $\gamma_1$ enters the oscillation equations. From Eq.\ (1) of
\citet{jcd:ADIPLS} we see that $\gamma_1$ only enters through terms similar to
the first term on the right-hand-side of Eq.\ (\ref{oscgm1})
\be{oscgm1}
    \frac{1}{\gamma_{1,{\rm tot}}}\dd{\ln p}{\ln r} \equiv
        \frac{1}{\gamma_1}\dd{\ln p_{\rm g}}{\ln r} +
        \frac{1}{\gamma_{1,{\rm turb}}}\dd{\ln p_{\rm turb}}{\ln r}\ ,
\ee
and we propose the total $\gamma_1$ to be given by the left-hand-side. Such a
$\gamma_{1,{\rm tot}}$ requires a knowledge of the total pressure, $p_{\rm g} +
p_{\rm turb}$. We can also define a total $\gamma_1$ that gives the response
in the total pressure, based on just the gas pressure
\be{gm1tld}
    \widetilde{\gamma}_{1,{\rm tot}} \equiv \dd{\ln p_{\rm g}}{\ln r}/
        \left[\frac{1}{\gamma_1}\dd{\ln p_{\rm g}}{\ln r} +
        \frac{1}{\gamma_{1,{\rm turb}}}\dd{\ln p_{\rm turb}}{\ln r}\right]\ .
\ee
We show this expression together with the pure gas $\gamma_1$ for our solar
simulation, in Fig.\ \ref{gm1tot_Sun}.
\mfigur{gm1tot_Sun}{7cm}
    {The total $\gamma_1$, including both thermo- and hydro-dynamic effects,
     for the solar simulation of our grid. The peak in
     $\widetilde{\gamma}_{1,{\rm tot}}$ reaches a value of 5.46.}
The reduced $\gamma_1$, adopted by \citet{rosenthal:conv-osc} as a limiting case
\be{gm1red}
    \gamma_{1,{\rm red}} = \gamma_1\frac{p_{\rm g}}
                            {p_{\rm g} + p_{\rm turb}}\ ,
\ee
is also shown in Fig.\ \ref{gm1tot_Sun}, and is rather different from our
Eq.\ (\ref{gm1tld}). \citet{rosenthal:conv-osc} found that using the gas
$\gamma_1$ for both pressures gave frequencies closer to the observed solar
ones, than using Eq.\ (\ref{gm1red}). With the rather different behavior of
Eq.\ (\ref{gm1tld}), there is a potential for improved agreement with
helioseismic observations, when using Eq.\ (\ref{gm1tld}).

This calculation is only a first step, since it assumed no phase-lag between
density and turbulent pressure, and since we assumed $\gamma_{1,{\rm turb}}$
to be independent of frequency. Future work with our grid
of simulations will address these very important issue, which has grown more
urgent with the advances in asteroseismology and the recent results by
\citet{savita:Kpl22sols}. They studied 22 solar-like stars, targeted by NASA's
Kepler mission, and found magnitudes of the surface effect that are not
immediately reconcilable with Fig.\ \ref{cExpan}, suggesting that
$\gamma_{1,{\rm turb}}$-effects are significant.

	\section{Variation of Granulation with Stellar Parameters}
	\label{varTg}

In this section we present the variation of granulation size and intensity
contrast with respect to the dependent variables of the grid, $T_{\rm eff}$
and $\log g$. In the case of our own star, the sizes of granules can be
measured directly from both ground- and space-based observations, and agree
well with simulations. The intensity contrast, on the other hand, is less
well constrained as observations are affected by both seeing, limited aperture,
diffraction and scattering from various parts of the telescope, focus issues,
etc., that can easily halve the observed contrast. Attempting to model these
effects in Hinode observations, \citet{danilovic:HinodeGranContr} found general
agreement with their MuRAM simulations, with which ours agree fairly well.
We find a 16.3\% RMS fluctuation in white light granulation for our solar
simulation, compared to their 14.4\% RMS contrast in 6\,300\,{\AA}
monochromatic light. \citet{beeck:3DsimComp} shows that the 5\,000\,{\AA}
contrast agrees between the simulations, although the detailed distribution
and granulation patterns do differ some.

It is worth noting that granulation in many ways are more constrained by
unresolved spectral observations \citep{nordlund:LivingRev}, as can also be
carried out for stars. This is due to the fact that convection in late-type
stars sets up correlations between temperature and velocity, which results in
C-shaped bisectors of emergent spectral lines
\citep{gray:FGK-LineBisecs,basturk:line-bisectors+3D}. The detailed line shapes
of our solar and Procyon simulations agree well with observations
\citep{asplund:solar-Fe-shapes,prieto:Procyon-conv-Fe}.

%
%
\subsection{Intensity Distribution of Granulation}
\label{env}

We have calculated 2D spatial power spectra of the white light, disk center,
surface intensity for each of the simulations, and averaged the spectra over
time (at least 10 periods of the fundamental p mode excited in each simulation).
These spectra are very broad, as can be seen from Fig.~\ref{int_spctr},
but the maxima, marked by filled circles and defining the typical size of
granulation, is never-the-less well-defined. The range of each spectrum
reflects the horizontal resolution at the lower end, and the horizontal extent
at the upper end. The similar relative location of the peak in each
spectrum means the granules are similarly well resolved.
\mfigur{int_spctr}{9.0cm}
	{The time averaged spatial spectra of emergent disk-center intensity
     for each of the simulations of the
	 grid. The thin lines are the raw spectra, and the thick lines are the
	 spectra smoothed with a Gaussian kernel with FWHM of 11 wavenumber points.
     We used short-dashed lines for giants ($\log g<2.1$), long-dashed lines
     for turn-off stars and solid lines for dwarfs ($\log g>4.1$). The solar
     simulation is shown in red.
     The abscissa, $d=2/(\sqrt{\pi}k)$, refers to diameters of circles with
     the same area as the $1/k^2$ square, where $k$ is the wavenumber
     corresponding to the Fourier transform.
     The maxima of the smoothed spectra are indicated with circles, and the
     corresponding granular sizes, $A_{\rm gran}$, are listed in table
     \ref{simtab}. The transformation of the maxima
     to the atmospheric HR-diagram, Fig.\ \ref{cAgran}, is bijective, with
     giants to the right and cool dwarfs at the lower left.}
This granular size, $A_{\rm gran}$,
is presented in Fig.~\ref{cAgran}. The granular size, $A_{\rm gran}$,
grows almost inversely 
proportional with the surface gravity and increases only slightly with
$T_{\rm eff}$. A least squares fit results in
\bea{AgranFit}
      \log\frac{A_{\rm gran}}{[{\rm Mm}]}
        &\simeq& (1.3210\pm 0.0038)\stimes\log T_{\rm eff} \mideqn
             &-& (1.0970\pm 0.0003)\stimes\log g\ + 0.0306\pm 0.0359\nonumber\ ,
\eea
where $g$ is in cgs-units and the $1\sigma$ uncertainties of the linear regression is indicated. So
the sensitivity to $T_{\rm eff}$ is slightly larger than to $\log g$, but
since the stellar range of $T_{\rm eff}$ is so much smaller than the range
in gravity, the gravity dependence is effectively the most important, as
also seen in Fig.~\ref{cAgran}. It is often speculated that the size of
granules scales with the pressure scale-height \citep{schwarzschild:RG-gran}
or the mixing length in the
atmosphere, but we find such factors ranging from $A_{\rm gran}/H_p\sim 9$--13
and $A_{\rm gran}/(\alpha H_p)\sim 5$--9, respectively, both generally
increasing from dwarfs to giants.
\mfigur{cAgran}{9.5cm}
	{The typical size of granulation (logarithmic scale) as function of
	 atmospheric parameters.
	 The granular size is found as the maxima of the smoothed spectra in
	 Fig.~\ref{int_spctr}.}

The time-averaged histograms of surface intensities has a bi-modal structure,
with most of the surface being covered in bright upflows with a narrower
distribution of intensities than the less bright, inter-granular lanes.

The intensity distributions of granules are narrower, due to the upflows being
nearly isentropic (See Sect.\ \ref{GenConv}). Their widths are determined by
the radiative losses suffered by the over-turning upflows at the top of the
convective envelope. If the top is located in optically deep layers, the
radiative losses well be relatively minor and both the width of the intensity
distribution of the granules, and the intensity contrast to the
inter-granular lanes are small---The granulation is largely hidden from our
view.

If the granulation reaches into optically thin layers, on the other hand,
the reverse is the case and we see both a broader distribution and a larger
contrast. This is often referred to as ``naked granulation'', a term coined
by \citet{stell-gran3}.

\begin{figure*}[hbt]
    \centerline{\psfig{figure=\figdir/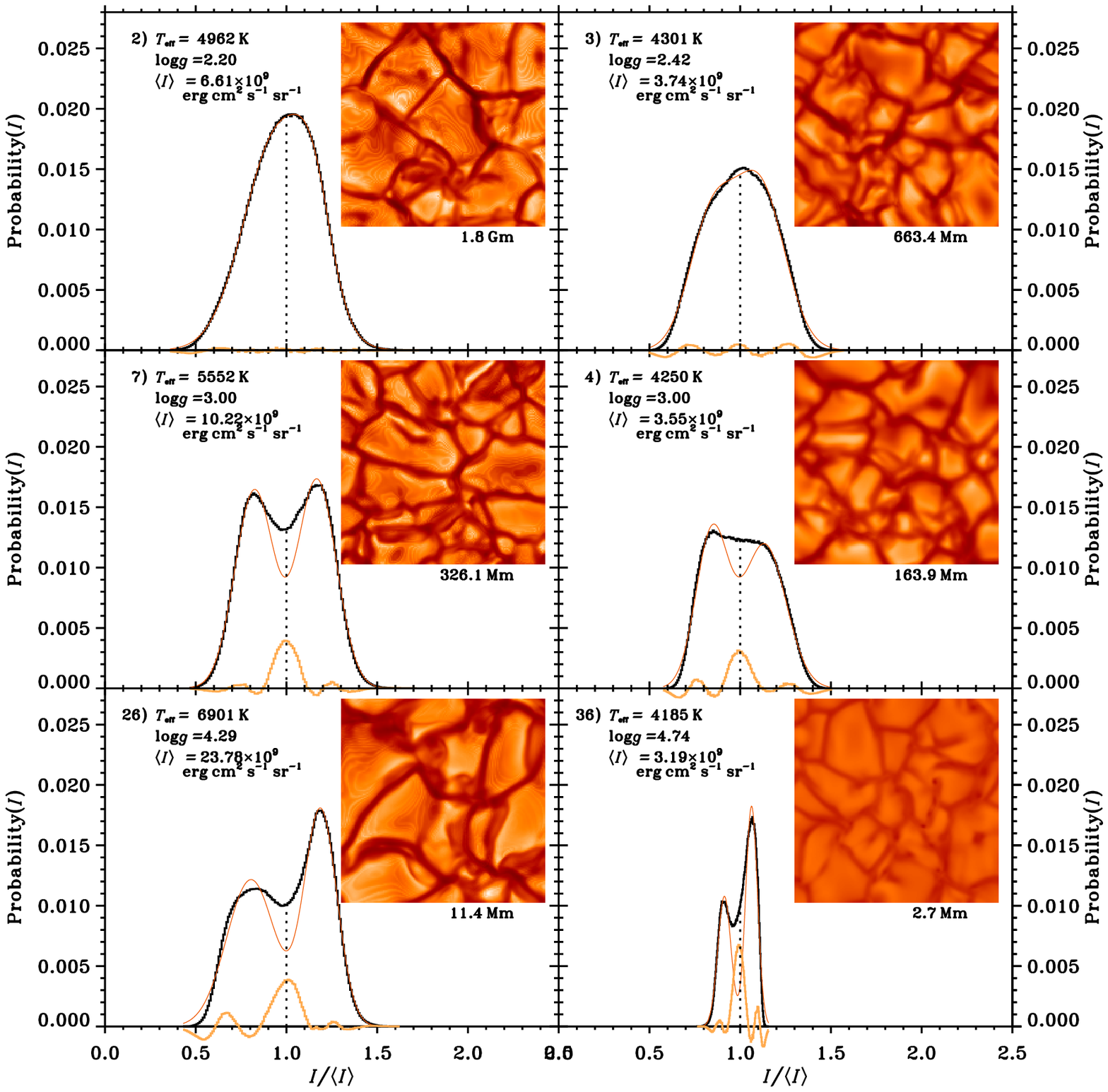,width=17.2cm}}
	\caption{\label{int_distr}
	The intensity distributions of each of the simulations of the grid
	(black histogram), together with a fit to Eq.~(\ref{IHistFit}) (thin orange
    line) and the difference between the two in yellow. From the differences we
    see a third component, increasing with gravity. The abscissa shows intensity
    normalized to the average for each simulation, to facilitate comparisons.
    This average intensity is listed together with the atmospheric parameters,
    in each panel. We also display
    white light, disk-center, surface intensity snapshots, with
    the horizontal extent of each simulation indicated below.
    The displayed snapshots were chosen to
    have intensity distributions close to the average for each simulation.}
\end{figure*}

The down-flows, on the other hand, are darker and have a much broader
distribution. These intensity distributions are well-approximated by a
%
%
\mfigur{cIRMS}{9.2cm}
	{As Fig.\ \ref{cSmax}, but showing the contrast of granulation,
     evaluated as the RMS scatter of the normalized intensity,
     $I/\langle I\rangle$.}
simple double Gaussian
\be{IHistFit}
	n(I) = I_1 e^{-((I-I_2)/I_3)^2} + I_4 e^{-((I-I_5)/I_6)^2}\ .
\ee
%
Fits to that expression are displayed in Fig.~\ref{int_distr}. Here we also
notice that the simulations display less of a dip between the two components,
compared to the double Gaussian fit. This third component, is due to the
breaking up of granules, which occurs in most of the snapshots shown here.
Some of the granules gets split apart by wedges of cool down-flows, others,
mainly the largest granules, have cool spots developing in their centers,
suddenly succumbing to the negative buoyancy and turning into a down-draft.
This occurs when the upflow in a granule (proportional to the area of the
granule) exceeds the capacity for horizontal outflow from the granule
(proportional to the circumference), braking the upflow and thus increasing the
cooling time. The developing cool spot, quickly connects to the nearest
inter-granular lane.
This phenomenon is also known as \emph{exploding granules}
\citep{namba:ObsExplGranules,spruit:SolarConv,nordlund+stein:RadDyn1991},
although
``exploding'' seems a bit dramatic for all, but the largest of these events.
\citep[See, {\eg}, ][for comparisons of simulations and observations of
exploding granules]{rast:expl-gran}.

The granulation in the snapshots of Fig.~\ref{int_distr} looks rather
different between the simulations, but this is a random occurrence, and the
morphology looks different for another set of snapshots of the same simulations
(cf.\ Fig.\ \ref{gran_form} for different snapshots of simulations 2, 26 and
37).
The over-all scale of the granulation is not apparent in Fig.~\ref{int_distr}
as the dimensions of each simulation is chosen so as to contain a similar number
of granules (about 30 granules),
and the only property that varies markedly and systematically
with atmospheric parameters, is the intensity contrast. We note that the
distribution appears bimodal for only the higher gravity simulations. For the
two giants of Fig.\ \ref{int_distr}, however, the distributions are not
symmetric and they are still best described by two separate Gaussians, as in
Eq.\ (\ref{IHistFit}). It is interesting that it is rather hard to pick out the
intensity distribution from the picture. One might recognize sharper edges in
the granules of simulation \#26, compared to that of \#4,
which gives rise to the separation of the two components in the distribution,
but it is not immediately obvious that, {\eg}, the two giants are similar.
We also note that the unfitted component from splitting or ``exploding''
granules, increases with surface gravity, to the point of being absent in the
simulations of giants. This behavior is not yet understood.

The contrast of the granulation, evaluated as the RMS scatter around the mean
is shown in Fig.~\ref{cIRMS}. From this figure we see how the contrast
behaves similarly to the convective efficiency, as measured by the atmospheric
entropy jump, Fig.\ \ref{cSjump}, with the highest contrast for the least
efficient convection. The slope of iso-contrast lines change significantly from
the cool to the warm side of the diagram, as opposed to the near-constant slopes
of the entropy jump. The granulation contrast increases rather linearly across
the contours, whereas the entropy jump increases close to exponentially, as
function of $\log T_{\rm eff}$ and $\log g$.

On the main sequence, for $\log g \ge 4$, a linear fit to the intensity
contrast gives
\bea{MSGranFit}
    I_{\rm RMS}/[\%]
             &=& (54.98\pm 1.86)\stimes\log T_{\rm eff} \mideqn
             &-&  (4.80\pm 0.53)\stimes\log g - 169.00\pm 0.73\nonumber\ ,
\eea
and towards the giants, for $\log g \le 3$,
\bea{GiantGranFit}
    I_{\rm RMS}/[\%]
             &=& (20.81\pm 1.34)\stimes\log T_{\rm eff}  \mideqn
             &-&  (1.11\pm 0.14)\stimes\log g -55.46\pm 0.29\nonumber\ .
\eea
Notice the change of $\log T_{\rm eff}$- and $\log g$-dependency between dwarfs
and giants,
recognized as the change of slope of the contours in Fig.\ \ref{cIRMS}. 
%
%
	\section{Conclusions}
	\label{conclusion}

We present an attempt at constructing a homogeneous, comprehensive,
grid of 3D convective atmosphere
simulations, with derived quantities for use by the general astronomical
community. The simulations are constructed to be as realistic as possible,
with realistic EOS, opacities and radiative transfer, in order to provide
a solid foundation for interpreting observations.

For all the subjects studied in the present paper, we found
significant differences with respect to conventional 1D stellar atmosphere
models---differences that will have an impact on the interpretation of most
stellar observations.

In Sect.\ \ref{GenConv} we gave an overview of the nature and morphology of
convection, as observed in the simulations of our grid. We see fast, turbulent,
entropy deficient downdrafts, driven by radiative cooling at the surface,
plowing through an ambient, isentropic, laminar upflow. This is rather different
from the picture normally associated with the MLT formulation, of distinct,
warm, rising convective elements, existing for a mere mixing-length.
We hope that the more realistic concept of convection, based on hydrodynamic
simulations, will provide an improved framework for further theoretical
developments in simplified, but realistic, descriptions of convection.

Components of the seismic surface effect, a systematic frequency shift of
observed p modes compared to 1D stellar model predictions, were explored in
Sect.\ \ref{SurfEff}. Part of the effect is due to enlarged acoustic cavities,
caused by a convective expansion of the atmosphere (with respect to 1D models).
This expansion depends smoothly on atmospheric parameters, ranging from
8.7\,km for the coolest dwarf in our grid, 140\,km for the Sun, to 170\,Mm for
the warmest giant.
We also attempt to derive an effective adiabatic gradient for the turbulent
pressure, which is another important component of the surface effect. This
$\gamma_{1,{\rm turb}}$ behaves rather differently from previous assumptions
about its properties. From the form of the adiabatic oscillation equations,
we propose the effective $\gamma_1$ should be a harmonic mean, weighted by
the pressure gradients, Eq.\ (\ref{oscgm1}).

Convection manifests itself as granulation at the surface, but has only been
directly observed for the Sun. In Sect.\ \ref{varTg}, we find that the size
depends smoothly on atmospheric parameters, but is not simply proportional to
atmospheric pressure scale-height, as has often been assumed for a lack of
observational constraints.

We have shown that employing realistic 3D convection simulations in the
interpretation of stellar observations will affect all stages of the analysis,
and also allow new questions to be asked---questions that have been beyond
simplified, 1D models of convection. In the near future our grid of simulations
will be used for calibrating the MLT mixing-length, and evaluate the excitation
and damping of p modes, among other applications.

\acknowledgments

The helpful comments and suggestions by the anonymous referee are much
appreciated.
We are grateful to W.\,D{\"a}ppen for access to the code and data tables
for the MHD equation of state.
RT acknowledges funding from the Australian Research Council (grants
DP\,0342613 and DP\,0558836) and NASA grants NNX08AI57G and NNX11AJ36G.
{\AA}N acknowledges current support from the Danish Natural Science
Research Council and from the Danish Center for Scientific Computing (DCSC).
RFS acknowledges NSF grant AGS-1141921 and NASA grant and NNX12AH49G.
This research has made extensive use of NASA's Astrophysics Data System.




\end{document}